# *In vitro* toxicity of nanoceria: effect of coating and stability in biofluids

N. Ould-Moussa[1], M. Safi[1], M.-A. Guedeau-Boudeville[1], D. Montero[2], H. Conjeaud[1] and J.-F. Berret[1*]

[1] *Matière et Systèmes Complexes, UMR 7057 CNRS Université Denis Diderot Paris-VII, Bâtiment Condorcet 10 rue Alice Domon et Léonie Duquet, 75205 Paris (France)*
[2] *Institut des Matériaux de Paris Centre (FR CNRS 2482), Université Pierre et Marie Curie, 75252 PARIS Cedex 05, France*

**Abstract:** Due to the increasing use of nanometric cerium oxide in applications, concerns about the toxicity of these particles have been raised and have resulted in a large number of investigations. We report here on the interactions between 7 nm anionically charged cerium oxide particles and living mammalian cells. By a modification of the particle coating including low-molecular weight ligands and polymers, two generic behaviors are compared: particles coated with citrate ions that precipitate in biofluids and particles coated with poly(acrylic acid) that are stable and remain nanometric. We find that nanoceria covered with both coating agents are taken up by mouse fibroblasts and localized into membrane-bound compartments. However, flow cytometry and electron microscopy reveal that as a result of their precipitation, citrate-coated particles interact more strongly with cells. At cerium concentration above 1 mM, only citrate-coated nanoceria (and not particles coated with poly(acrylic acid)) display toxicity and moderate genotoxicity. The results demonstrate that the control of the surface chemistry of the particles and its ability to prevent aggregation can affect the toxicity of nanomaterials.

*Corresponding author. Email: jean-francois.berret@univ-paris-diderot.fr

# Introduction

Due to their unique size-dependent catalytic and optical properties, nanometric cerium oxide ($CeO_2$) particles have found applications in the automotive industry, in chemical-mechanical polishing and in paint formulation. Concerns about their toxicity have been raised however and have resulted in a large number of studies recently (HEI 2001; Cassee et al. 2011; Karakoti et al. 2010; Walser et al. 2012; Horie et al. 2012). During the last decade, the uptake and toxicity of $CeO_2$ nanoparticles were investigated on both prokaryotic (Thill et al. 2006) and eukaryotic (Limbach et al. 2005a; Lin et al. 2006; Das et al. 2007; Brunner et al. 2006; Xia et al. 2008b; Park et al. 2008; Safi et al. 2010; Horie et al. 2011; Chigurupati et al. 2013; Culcasi et al. 2012; Lee et al. 2012; Simonelli et al. 2011; Kroll et al. 2011) cells. *In vitro* assays were performed on various cell lines, including lung epithelial and cancer cells, macrophages using particles obtained from a wide variety of synthesis. Most reports revealed that nanoceria were internalized into cells (Limbach et al. 2005a; Xia et al. 2008b; Yokel et al. 2009; Horie et al. 2011; Hussain et al. 2012a; Ma et al. 2012). With regards to the toxicity and to the emission of reactive oxygen species (ROS), results from the literature are controversial. Some authors demonstrated the induction of acute toxicity (Thill et al. 2006; Brunner et al. 2006; Auffan et al. 2009; Eom and Choi 2009; Cho et al. 2010; Ma et al. 2012; Srinivas et al. 2011; Zhang et al. 2011; Hussain et al. 2012a) and oxidative stress (Lin et al. 2006; Park et al. 2008; Horie et al. 2011), while others provide evidences that the nanoparticles were non-toxic (Xia et al. 2008b; Safi et al. 2010; Birbaum et al. 2010; Pierscionek et al. 2012; Lee et al. 2012). In some instances, the particles exhibited strong anti-oxidant properties that promote cell survival under oxidative stress conditions (Das et al. 2007; Hirst et al. 2009; Xia et al. 2008b; Karakoti et al. 2010). These contradictory results could have several grounds, such as the synthesis of the particles (Karakoti et al. 2010), the cell lines investigated or the incubation protocols.

On the physico-chemical side, it has been realized that the interactions between living cells and particles *in vitro* depend eventually on the behavior of these particles in biological fluids. The two main scenarios noted in the literature are the formation of a protein corona (Rocker et al. 2009; Casals et al. 2010; Walczyk et al. 2010; Sund et al. 2011; Safi et al. 2011) and the aggregation into micro-size clusters (Williams et al. 2006; Diaz et al. 2008; Petri-Fink et al. 2008; Chanteau et al. 2009; Safi et al. 2010; Safi et al. 2011). In most studies on the risk assessment of nanoceria, the particles used were not coated, and therefore precipitated in cell preparation media (Lin et al. 2006; Xia et al. 2008b; Auffan et al. 2009; Chanteau et al. 2009; Vincent et al. 2009; Eom and Choi 2009; Zhang et al. 2011; Hussain et al. 2012b; Yokel et al. 2009; Ma et al. 2012). When aggregated, the hydrodynamic properties of the nanoparticles are strongly modified, and so are their interactions with cells. Reports mentioned that to avoid precipitation, the concentrations had to be kept very low, typically below 10 μg mL$^{-1}$ (Lin et al. 2006; Eom and Choi 2009). Others studies used controlled adsorption processes of organics at the water/oxide interfaces, including oligomers (phosphonate-terminated poly(ethylene oxide) (Chanteau et al. 2009)), proteins from fetal bovine serum (Horie et al. 2011) or polymers (poly(acrylic acid) (Safi et al. 2010), poly(acrylic acid)-*b*-poly(acrylamide) (Galimard et al. 2012)). The later processes ensured an excellent stability in biofluids for long periods of time (> weeks).

In the present paper, we investigate the interactions between 7 nm anionically charged cerium oxide particles and embryonic fibroblast cells (NIH/3T3). The coating of the particles was provided by low molecular weight ligands (citric acid) and by ion-containing polymers (poly(acrylic acid)), both being carboxylate terminated. Doing so, two generic behaviors were compared: citrate-coated particles that precipitate in biofluids and polymer-coated particles that are stable and remain nanometric. We have found that nanoceria are indeed uptaken by cells and localized into membrane-bound compartments. At concentrations used in this study, the particles displayed no acute short-term toxicity, nor they induce oxidative stress. At high concentration only, aggregating citrate-coated nanoceria display toxicity, moderate ROS expression and genotoxicity.

# Materials and Methods

## Chemicals, synthesis and cell culture

*Nanoparticles synthesis and coating*: Cerium oxide nanoparticles were synthesized by thermo-hydrolysis of cerium nitrate salt under hydrothermal conditions at neutral or acidic pH. Provided to us by Solvay (Centre de Recherche d'Aubervilliers, Aubervilliers, France), the dispersion arises from the same type of synthesis as EOLYS®, a Solvay product traded in as fuel-borne catalyst for diesel cars. Here however and in contrast to EOLYS, the particles are dispersed in a water-borne solvent. Synthesized at acidic pH (pH 1.5), the particles are cationic nanocrystals with nitrate counterions adsorbed at their surfaces (S1-S2). Electrostatics ensures the stabilization of the dispersion. Increases in pH or ionic strength induced the destabilization of the sols and the irreversible aggregation of nanoceria (Nabavi et al. 1993). Transmission electron microscopy (TEM) performed on a dilute dispersion exhibits isotropic agglomerates made of 2 nm crystallites with faceted morphologies (Fig. 1). The size distribution of the agglomerates obtained by TEM was adjusted by a log-normal function with median diameter 7.0 nm and polydispersity 0.15. The polydispersity is defined as the ratio between the standard deviation and the average diameter.

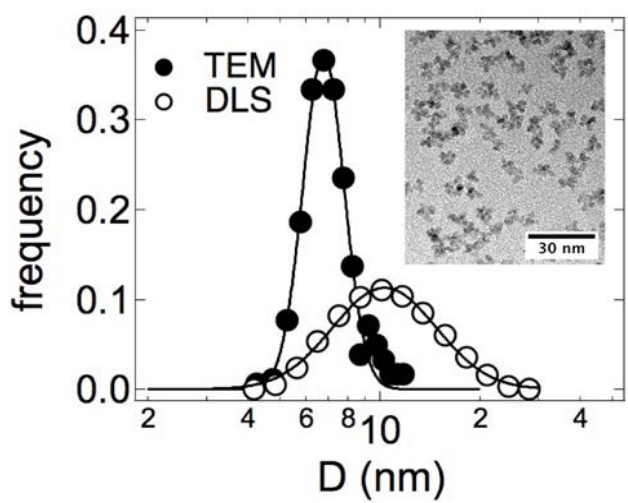

***Figure 1:*** *Size distributions of the bare nanoceria ($CeO_2$) determined by transmission electron microscopy (number distribution, closed circles) and dynamic light scattering (intensity distribution, empty circles). The median diameters are 7.0 and 9.8 nm respectively. Both distributions are normalized so that their sum is 1. Inset: electron microscopy image of the particles (magnification 120000×).*

Using static (SLS) and dynamical (DLS) light scattering, the hydrodynamic diameter $D_H$ and the molecular weight $M_W$ of the particles were found at $D_H = 9.8$ nm and $M_W = 3.3 \times 10^5$ g $mol^{-1}$, respectively (Qi et al. 2008b; Qi et al. 2008a). In Fig. 1, the distribution in hydrodynamic diameters is slightly shifted with respect to that of TEM, the reason being that when particles are distributed, light scattering is sensitive to the largest particles.

In this work, two types of coating were used: citric acid and poly(acrylic acid). Uncoated nanoparticles were not considered because of their pronounced instability in physiological media. Citric acid (molecular weight 192.1 g $mol^{-1}$, Aldrich), is a weak triacid with $pK_a$ at 3.1, 4.8 and 6.4. Performed after the synthesis, the ionization of the carboxyl groups has the effect to reverse the surface charge of the particles, which hence became negatively charged. The charge density was estimated at $-2e$ $nm^{-2}$ (Berret 2007). With such low molecular-weight ligands, the adsorbed citrates are in equilibrium with free citrate molecules dispersed in the bulk solution, reflecting an adsorption isotherm behavior. Cit-CeO$_2$ nanoparticles exhibit similar hydrodynamic sizes as the bare particles (Chanteau et al. 2009). Poly(acrylic acid) was frequently used in the past for coating inorganic particles (Qi et al. 2008b; Qi et al. 2008a). Here, we used poly(sodium acrylate) with molecular weight of 2100 g $mol^{-1}$ (polydispersity 1.7). The coating of the nanoparticles was performed according to the precipitation-redispersion protocol (Berret et al. 2007). The adsorbed polymer corona was estimated at ~ 3 nm by means of dynamical light scattering, yielding a $D_H$ of 15 nm. The number of adsorbed chains per particle was 57 (Qi et al. 2008a), corresponding to a density of chargeable groups of 4 $nm^{-2}$. For the cell culture assays, the dispersions were dialyzed against deionized water using the biocompatible Spectra Por 2 membrane of MWCO 12 kD. For the dialysis of the citrate coated particles, the dialysis bath was supplemented with 10 mM of citrates.

*NIH/3T3 cell culture:* The NIH/3T3 embryonic fibroblast is a standard cell line already tested with many particles in the context of toxicity studies (Hillaireau and Couvreur 2009; Cassee et al. 2011; Karakoti et al. 2012). Fibroblasts are moreover the most common cells of connective tissues in animals and represent the first barrier to toxic substances. In nanoceria-based applications such as chemical and mechanical polishing or paint formulation, the particles can come into contact with the skin and interact with fibroblasts. Several recent studies have reported the use of fibroblasts in the assessment of the toxicity of engineered nanoceria (Chigurupati et al. 2013; Culcasi et al. 2012; Lee et al. 2012; Simonelli et al. 2011; Kroll et al. 2011). The fibroblast cells were grown as a monolayer in Dulbecco's modified Eagle's medium (DMEM) with high glucose (4.5 g.l$^{-1}$) and stable glutamine (PAA Laboratories GmbH, Austria). This medium was supplemented with 10% fetal bovine serum (FBS) and 1% penicillin/streptomycin (PAA Laboratories GmbH, Austria), referred to as cell culture medium. Exponentially growing cultures were maintained in a humidified atmosphere of 5% CO$_2$ – 95% air at 37 °C, and under these conditions the plating efficiency was 70–90% and the doubling time was 12 – 14 h. The cell cultures were passaged twice a week using trypsin–EDTA (PAA Laboratories GmbH, Austria) to detach the cells from their culture flasks or wells followed by the addition of cell culture medium to neutralize the trypsin. The cells were pelleted by centrifugation at 1200 rpm for 5 min at 25 °C. The supernatant was removed and cell pellets were re-suspended in assay medium and counted using a Malassez counting chamber. Electron microscopy, flow cytometry and comet assays were performed on cells incubated with citrate and poly(acrylic acid) coated nanoceria at 0.1, 1 and 10 mM of cerium concentration. These concentrations correspond to 0.0017, 0.017 and 0.17 wt. %, or 17, 170 and 1700 µg mL$^{-1}$ respectively, and are representative of those reported in the literature for *in vivo* and *in vitro* assays on a wide variety of cells (Karakoti et al. 2010; Thanh and Green 2010; Cassee et al. 2011; Iversen et al. 2011).

## Experimental techniques

*UV-visible spectroscopy* : A UV–visible spectrometer (Cary 50 Scan from Varian) was used to measure the absorbance of bare and coated nanoparticle dispersions in water. In the range $\lambda$ = 200 – 1100 nm, the absorbance was related to the concentration by the Beer–Lambert law: $Abs(c, \lambda) = \varepsilon(\lambda)c\ell$, where $\ell$ (=1 cm) is the optical path length, $c$ the nanoparticle concentration and $\varepsilon$ the molar absorption coefficient (cm$^{-1}$ M$^{-1}$). Taking advantage of the strong absorbance of cerium oxide below 400 nm (S3), the cerium concentrations could be determined accurately, with an uncertainty better than $5\times10^{-4}$ wt.% (5 µg mL$^{-1}$) (Safi et al. 2010). Citrate and polymer coating did not modify the absorption coefficient.

*Dynamic light scattering* : Dynamic light scattering was performed on a Brookhaven spectrometer (BI-9000AT autocorrelator, $\lambda$ = 632.8 nm) for measurements of the scattered intensity $I_S$ and the collective diffusion constant $D_0$. In dynamic light scattering, the collective diffusion coefficient $D_0$ was determined from the second order autocorrelation function of the scattered light. From the value of the coefficient, the hydrodynamic diameter of the colloids was calculated according to the Stokes–Einstein relation, $D_H = k_B T / 3\pi\eta_S D_0$, where $k_B$ is the Boltzmann constant, $T$ the temperature (T = 298 K) and $\eta_0$ the solvent viscosity ($\eta_S = 0.89 \times 10^{-3}$ Pa s for water at T = 25 °C). The autocorrelation functions were interpreted using the cumulants and the CONTIN fitting procedure (intensity distribution) provided by the instrumental software. To study the kinetics of aggregation, light scattering experiment was performed using the same protocol as for incubation on a Zetasizer Nano-ZS (Malvern Instruments Ltd., UK). A test tube containing 1 ml of the cellular medium (DMEM) was placed on the spectrometer and a small volume (~ 10 µL) of a concentrated dispersion was poured rapidly in the tube and the sample was homogenized. The final concentration was [Ce] = 1 mM, corresponding to c = $1.72\times10^{-2}$ wt. % (172 µg mL$^{-1}$). The scattering intensity $I_S$ and the hydrodynamic diameter $D_H$ were monitored as a function of the time for 2 hours after injection, a time that was considered to be sufficient to conclude about the nanoparticle behavior in the solvent (Chanteau et al. 2009). In the dilute regime, the scattering intensity scales as $I_S = K M_W c$, where $K$ is the scattering contrast, $M_W$ and $c$ the molecular weight and concentration of the scatterers. If the particles aggregate, both $I_S$ and $D_H$ increase due to the augmentation of size and molecular weight of the scatterers. If the dispersion is stable, no variation is observed. For experiments performed with the cell culture medium, the particle concentration was raised to c = 0.172 wt. % (10 mM). In these conditions, the scattered intensity came predominantly from the particles, and not from the proteins present in the solvent.

*Optical microscopy:* Phase-contrast images of the cells were acquired on an IX71 inverted microscope (Olympus) equipped with 10× and 60× objectives. Data acquisition and treatment were monitored with a Photometrics Cascade camera (Roper Scientific) and treated with Metaview (Universal Imaging Inc.) and ImageJ softwares.

*Transmission electron microscopy (TEM) :* Electron microscopy measurements on the nanoparticles were carried out on a JEOL-100 CX microscope at the SIARE facility of the Université Pierre et Marie Curie (Paris 6). The TEM images of the cerium oxide nanoparticles (magnification 120 000×) were analyzed using ImageJ software (http://rsb.info.nih.gov /ij/). The diameter and polydispersity for $CeO_2$ were in good accordance with those determined by cryogenic transmission electron microscopy in an earlier report (Qi et al. 2008a). For TEM experiments on living cells, the 3T3/NIH fibroblasts were first seeded onto a 6-well plate and incubated with the nanoparticles during 24 h. Excess medium was then removed, and the cells were washed in 0.2 M phosphate buffer (PBS), pH 7.4 and fixed in 2% glutaraldehyde-phosphate buffer 0.1 M for 1 h at room temperature. Fixed cells were washed in 0.2 M PBS. After washing, samples were post-fixed for 45 minutes in a 1% osmium-phosphate buffer 0.1 M at room temperature in dark conditions. The samples were then dehydrated in reagent-grade ethanol. Samples were infiltrated in 1:1 ethanol/epon resin for 1 h and finally in 100% epon resin for 48 h at 60 °C for polymerization. 90 nm-thick sections were cut with an ultramicrotome (LEICA, Ultracut UCT) and picked up on copper-rhodium grids. They were then stained for 7 min in 2% uranyl acetate and for 7 min in 0.2% lead citrate. Grids were analyzed with a transmission electron microscope (ZEISS, EM 912 OMEGA) equipped with a $LaB_6$ filament, at 80 kV and images were captured with a digital camera (SS-CCD, Proscan 1024×1024), and the iTEM software.

*Scanning Electron Microscopy (SEM):* Cells were primarily fixed 1 h at room temperature and overnight at 4 °C by immersion in a fixative solution (2.5% glutaraldehyde in 0.1 M sodium cacodylate buffer, pH 7.5), washed 3 times with cacodylate buffer 0.2 M, and post fixed during 1 h at room temperature in 1% osmium tetroxide in 0.1 M sodium cacodylate buffer (pH 7.5) and washed again three times in 0.1 M cacodylate buffer. Dehydration until 100% ethanol was completed through graded ethanol-water mixtures at room temperature. Cells were then dried according to the $CO_2$ supercritical drying process (using a Bal-Tec CPD030). After mounting onto scanning stubs, samples were coated with a 10 nm conductive carbon layer using a thermal carbon evaporator (Cressington C208). Scanning electron microscopy was performed either on a Zeiss ULTRA 40 or a Hitachi SU-70 field emission scanning electron microscopes. Microanalysis and mapping were performed either on Edax CDU or Oxford X-Max EDX detectors installed on the microscope columns (S7).

*Flow cytometry: interactions and viability :* Coated nanoceria at concentrations 0.1, 1 and 10 mM were add to NIH/3T3 cells culture in complete DMEM medium and incubated for 5 h and 24 h. To evaluate the effect of the coating, sodium citrate and poly(sodium acrylate) were added separately to the fibroblasts. Concentrations ranging from 0.1 to 10 mM for the low-molecular weight ligand and from 0.01 to 0.1 wt. % for the polymer were used. These later values correspond to roughly 1 and 10 mM of carboxylate monomers. The cells were kept in normal culture conditions at 37 °C in a humidified atmosphere of 5% $CO_2$ and 95% air. The cells were detached using trypsin–EDTA (PAA Laboratories GmbH, Austria) and complete cell culture medium was added to neutralize the trypsin. The cells were then pelleted by centrifugation at 1200 rpm for 5 min at 25 °C. Finally, supernatants were removed and cell pellets were re-suspended in culture medium. For viability assays, propidium iodide (PI, Sigma Aldrich) was added to the cell suspension at a concentration of $2\mu g\ mL^{-1}$ five minutes before the cytometric analysis and performed in triplicate for particles as well as coating agents. Forward scatter (FS), side scatter (SS) and propidium iodide fluorescence of individual cells incubated in the conditions described previously were measured on an EPICS 4C flow cytometer (Beckman Coulter). Forward scatter FS and SS intensities (measured at 488 nm) were digitalized on both linear and logarithmic scale (4 decades). Red fluorescence intensities emitted by dead cells (measured at 610 nm) were digitalized on a logarithmic scale. Cellular debris and nanoparticles aggregates were removed by an appropriate choice of the gate in the cytogram FS versus LogSS (GATE1 in Fig. 2a). Red fluorescence (Fig. 2b) was used to define PI permeant cells (GATE3 for cells in M1, red dots) and PI non-permeant cells (GATE4 for cells not in M1, green dots). Dead and living cells were identified thanks to their scattering properties (Fig. 2a). Living cells were defined as cells present in GATE2 and GATE4 whereas dead cells were defined as cells present in GATE1 and GATE3. Finally, the SS (logarithmic scale in Figs. 2c and 2d) and FS (linear scale in Figs. 2e and 2f) intensities were reported for the living (Figs. 2c and 2e) and dead cells (Figs. 2d and 2f). The different parameters were visualized and median values calculated in each experimental condition.

*Flow cytometry: Reactive Oxygen Species :* Cellular generation of reactive oxygen species (ROS) was determined using dihydroethidium (Sigma Aldrich) as fluorescent biomarkers. Dihydroethidium (DHE) is a membrane-permeable dye which forms ethidium in the presence of the superoxide $O_2^-$. The resulting ethidium intercalates with DNA and provides an increase of its fluorescence at 620 nm upon excitation at 488 nm. NIH/3T3 cells were treated with citric acid ligands, with $PAA_{2K}$ polymers and with ceria nanoparticles in conditions similar to those of the toxicity assays. Pyocyanin (PCN, Sigma Aldrich) was used as a positive control for the ROS generation (Horke et al. 2010). PCN solution (20% DMSO) was diluted to a concentration of $50\mu M$ in the DMEM medium prior to incubation. After treatment, wells were washed three times with phosphate-buffered saline. The cells were further incubated with a DMEM medium containing 2 μM DHE for 30 minutes at 37°C in a humidified atmosphere of 5% $CO_2$ and 95% air. Cells were washed again with PBS to remove the DHE excess, trypsinized and dispersed in Phenol Red-free DMEM medium. Cytometric data were obtained on the EPICS 4C flow cytometer where the ethidium fluorescence was measured at 620 nm after excitation at 488 nm. Living cells were identified thanks to their scatter properties (similar to GATE2 in the viability analysis) and defined on the FS *versus* SS cytogram. The mean values of the red fluorescence intensities (digitalized on a logarithmic scale) were used as a measure of the intracellular ROS activity.

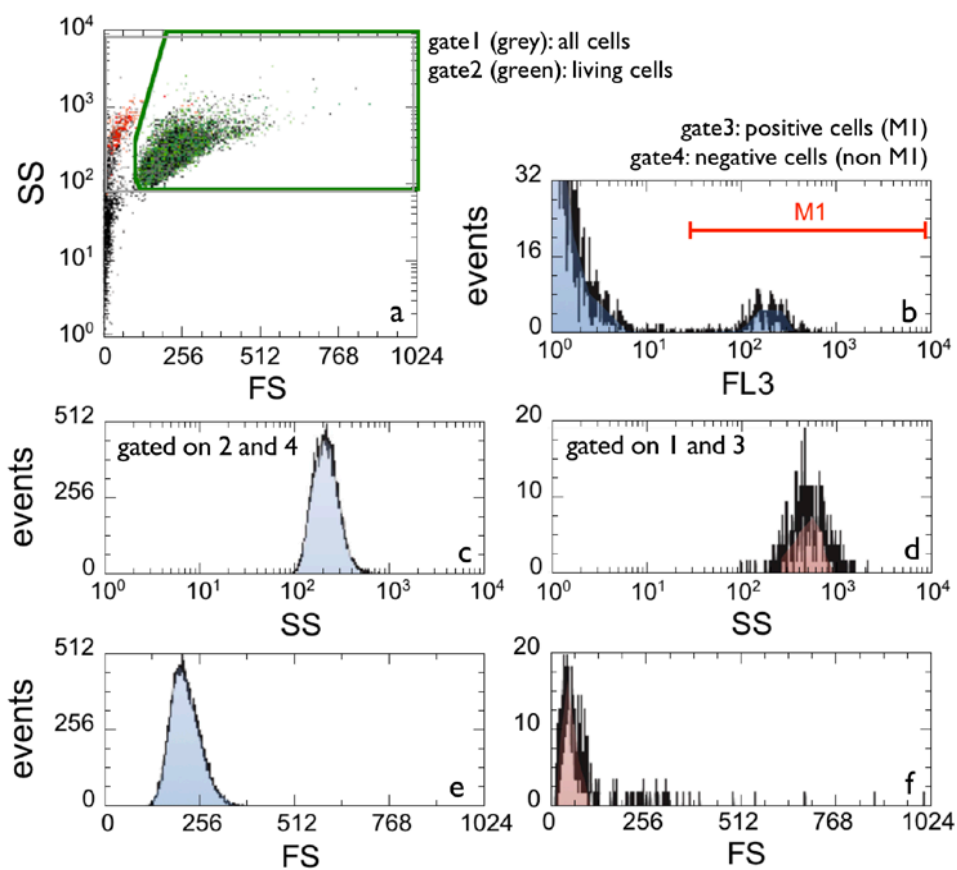

***Figure 2***: *Cytometric analysis: **a)** Dot plots of the side scatter (SS) intensities versus forward scatter (FS) intensities in semi-logarithmic scale. Cellular debris and nanoparticles aggregates were removed by an appropriate choice of the GATE1. b) Red fluorescence was used to define propidium iodide (PI) permeant cells (GATE3 for cells in M1, red dots) and PI non-permeant cells (GATE4 for cells not in M1, green dots). Living cells were defined as cells present in GATE2 and GATE4 whereas dead cells were defined as cells present in GATE1 and GATE3. c), d), e) and f) The SS (logarithmic scale in c and d) and FS (linear scale in e and f) intensities were reported for the living (c and e) and dead cells (d and f).*

*Comet assay*: The single-cell gel electrophoresis comet measurements were performed in alkaline condition. Cells were marked with Cit-CeO$_2$ nanoparticles and incubated for 4 hours. The incubation and the detachment of the cells were performed as described above in the case of the cytometry assay. The obtained cell suspensions after detachment were mixed at 37°C with low melting agarose solution (1% - Sigma Aldrich) in order to obtain a final suspension with a concentration of agarose 0.5 wt.%, this suspension was then pipetted onto a microscope slide which was previously coated with a thin deposit of agarose. All following steps were performed under dim light to prevent the occurrence of additional DNA damage. After incubating the slides for 5 minutes at 4°C, the slides were transferred for 1h in a lysis solution (*NaCl 2.5M, (EDTA)Na₂ 0.1M, Tris 0.01M, N-Sarcosinate 1%*, the lysis solution was previously adjusted to pH 10 by adding *NaOH* 10M). In order to unwind the DNA, the slides were plunged at room temperature for 40 minutes into alkaline buffer (a mixture of *NaOH* 0.3 M and *(EDTA)Na₂* $10^{-3}$M), the electrophoresis was then done for 25 min at room temperature with 23 V and adjusted to 300 mA by lowering the buffer level in the electrophoresis chamber. After electrophoresis, the slides were rinsed three times with a Tris buffer solution (0.4 M at pH 7.5), the DNA of each slide was colored with the addition of 75 µl of ethidium bromide (GIBCO-BRL) at a concentration of 20 µg mL$^{-1}$ in water. Finally, the slides were covered with a coverslip and stocked in the dark at 4°C in humid atmosphere. The DNA comets were observed with a fluorescent microscope (Orthoplan, Leitz, Germany). We used the software Comet Assay 2 (*Perceptive Instrument*) in order to analyze the measured comets, the product of the displacement between the tail of the comet and the head of the comet with the percentage of DNA in the tail was defined as the tail moment.

# Results

## Colloidal stability in physiological media

Light scattering experiment was performed using the same protocol as for the incubation of fibroblasts by nanoparticles. Few microliters of a concentrated nanoceria dispersion were poured and homogenized rapidly in the tube containing a physiological medium at time t = 0. Light scattering was hence monitored *versus* time. Figs. 3a and 3b display the hydrodynamic diameter $D_H$ for PAA$_{2K}$–CeO$_2$ and Cit–CeO$_2$ dispersed in phosphate buffered saline (PBS1x) and in a complete cell culture medium (DMEM), respectively. The Cit–CeO$_2$ nanoparticles are found to destabilize, as in both cases the diameter increases with time. The aggregation is slow in PBS1x and much more rapid in DMEM. In the supporting information section (S4), it is shown that the destabilization in PBS1x is completed after 8 hours. The slow decrease of the diameter observed with DMEM after 20 min is related to the sedimentation of the largest agglomerates (Fig. 3b). For PAA$_{2K}$-CeO$_2$ in contrast, $D_H(t)$ remains flat, the dispersion showing no sign of agglomeration whatsoever, even after 24 h. The $D_H$ of PAA$_{2K}$–CeO$_2$ particles passes from 15.2 nm in PBS1x to 16.5 nm in complete DMEM, suggesting that the particles are also devoid of a protein corona. Studied by light scattering, the complete cell culture medium DMEM exhibits broad size distributions of scatterers centered around 10 - 20 nm and extending to diameters as high as 100 nm. In DMEM, the scatterers are proteins, aggregates of proteins or biomacromolecules present in the formulation. If proteins or biological molecules were adsorbed on the external layer of the particles, the size of the complexes should be larger than at least 25 nm (S5). The images of the dispersions on the right hand side of the figure (Figs. 3c-f) were taken at 24 h and confirm the short time measurements: precipitation for the citrate coated particles and stability for the polymer coated particles. Similar results were found for iron oxide (γ-Fe$_2$O$_3$) nanoparticles coated with citrates, and the precipitation in biological fluids was attributed to the displacement of the ligands from the particle surfaces towards the bulk as they are preferentially complexed by counterions of the culture medium (Safi et al. 2011).

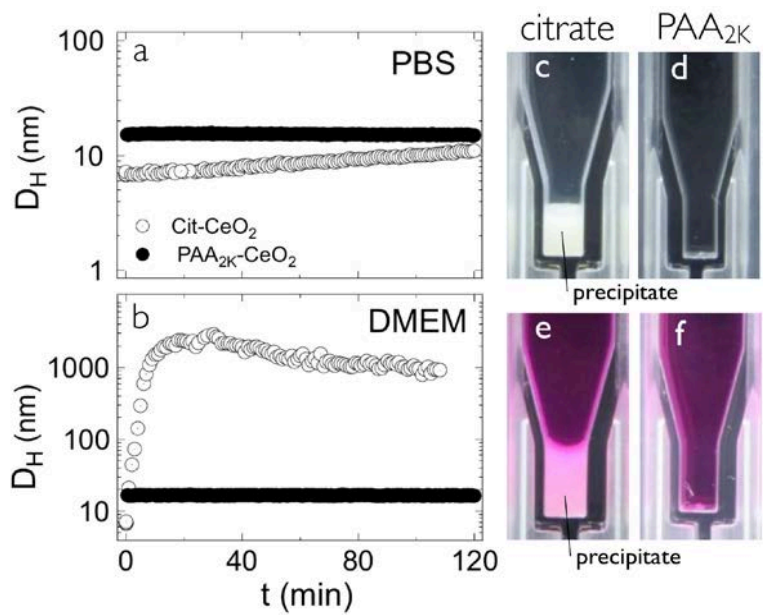

***Figure 3:*** *Hydrodynamic diameter of $PAA_{2K}$–$CeO_2$ and $Cit$-$CeO_2$ particles in PBS buffer (a) and Dulbecco's modified Eagle's medium supplemented with 10% fetal bovine serum (b). In these experiments, the cerium concentration was 10 mM (0.172 wt. %) and the temperature 37 °C. The pictures (c, d, e and f) in the right-hand side panels show the dispersions 24 h after mixing. In both media, $PAA_{2K}$-coated particles remained stable (d, f). Citrate coated nanoceria in contrast precipitated and settled down at the bottom of the cell (c, e).*

## Interaction cell/nanoparticle and localization

To monitor the interactions between the nanoparticles and the cells, each step of the incubation process was sequentially analyzed with different techniques: *i)* optical microscopy to visualize the treated fibroblasts in Petri dish environments, *ii)* scanning electron microscopy coupled to EDX to identify particles adsorbed at the plasma membrane and *iii)* electron transmission electron to track the internalized proportion.

*Phase-Contrast Optical Microscopy:* Figs. 4 show phase contrast images of NIH/3T3 cells obtained with optical microscopy (objective 60×). In Figs. 4a and 4b, the fibroblasts were exposed to respectively $PAA_{2K}$–$CeO_2$ and to Cit–$CeO_2$ at [Ce] = 1 mM (c = 0.017 wt. % or 170 μg mL$^{-1}$). The data are compared to the control in Fig. 4c. At this concentration, the cells exhibit normal and comparable growth rates (Safi et al. 2010). With citrate-coated particles, micrometric aggregates (arrows) sedimented on the bottom side of the Petri dish or on the cell membrane are clearly visible, confirming the agglomeration seen in the previous section.

*Scanning Electron Microscopy:* Scanning Electron Microscopy (SEM) was performed to visualize the particles at the plasma membrane of treated cells. The experimental conditions were an incubation time of 2 h and a CeO$_2$ concentration [Ce] = 1 mM (c = 0.017 wt. % or 170 μg mL$^{-1}$).

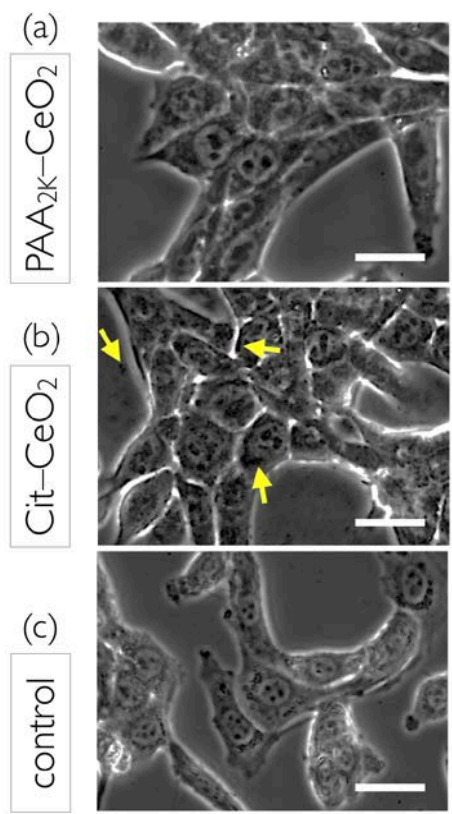

**Figure 4:** *60×-phase contrast microscopy images of NIH/3T3 fibroblasts incubated with 1 mM poly(acrylic acid) (a) and citrate (b) coated nanoceria at 24 h. Arrows indicate aggregates of citrate coated particles. The data are compared to a control (c). The bar is 20 μm.*

Figs. 5a and 5b shows images of a control cell. The cylindrical body of the NIH/3T3 exhibits at its surface protrusions identified as microvilli and indicated by red arrows (Eugene et al. 2002; Lorenz et al. 2006; Kemp et al. 2008; Zhang et al. 2010; Koeneman et al. 2010). Starting at the level of the membrane, these extensions are involved in several cellular functions including internalization, adhesion and mechano-transduction. For NIH/3T3 fibroblasts, the average length, diameter and density of the microvilli are respectively 600 nm, 160 nm and 3 μm$^{-2}$ (S6), in good agreement with literature data (Hecht et al. 2011). Figs. 5c and 5d display representative images of the plasma membrane for a cell incubated with PAA$_{2K}$–CeO$_2$. The cell body and the microvilli covering the cell surface maintained their morphology after a 2 h exposure. With the polymer coating, the nanoceria were difficult to localize, and very few could be detected (Fig. 5d). Cells treated with PAA$_{2K}$–CeO$_2$ indeed resemble the control. The surfaces of fibroblasts incubated with citrate-coated particles appeared differently (Fig. 5e, 5f). There, single nanoceria particles and particles clusters are deposited on the bilayer, or interact with the microvilli, forming with them complex and disordered sub-structures (Fig. 5f). The aggregate sizes at the cellular membrane were found in the range 20 – 500 nm, a result that was also found with iron oxide particles (Galimard et al. 2012). Note that due to the deposition of a 10 nm conductive carbon layer on the cell samples, the particles at the cellular membrane appear larger than they actually are ($\sim$ 20 nm). Energy dispersive X-ray analysis confirms that the particles present at the surface of the cells contained cerium atoms, their proportion being however in the detection limit of the technique, i.e. of the order of 1% (S7). In conclusion, SEM studies demonstrate the importance of visualization for distribution of resilient particles at the cellular level.

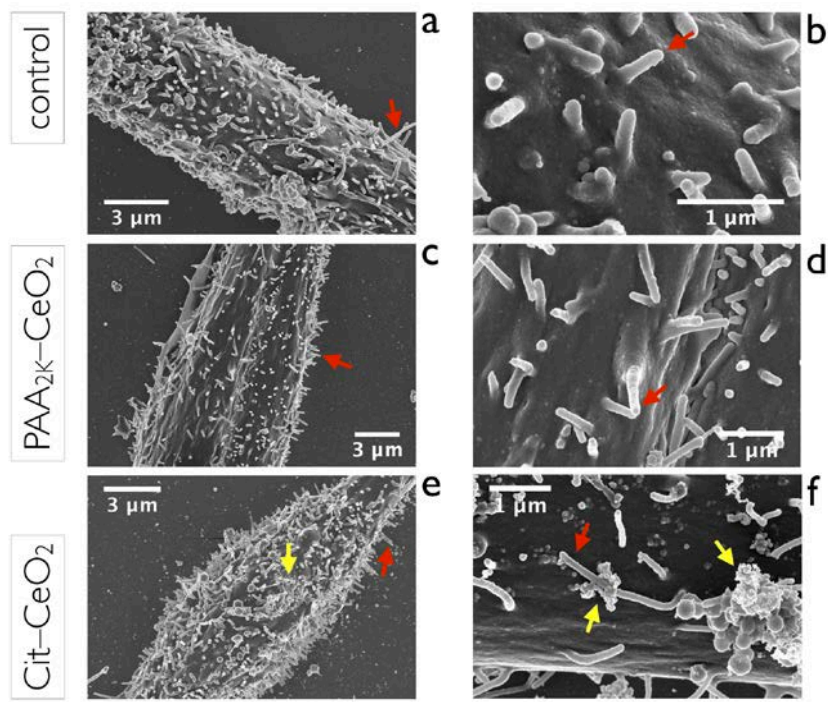

**Figure 5:** *Scanning Electron Microscopy images of NIH/3T3 untreated cells (a,b) and of cells treated with $PAA_{2K}$–$Ce_2O$ (c,d) and with Cit–$Ce_2O$ (e,f). Incubation time was 2 h and the cerium molar concentration in the supernatant [Ce] = 1 mM. The protrusions at the cell surface were identified as microvilli (red arrows). With the $PAA_{2K}$ coating, particles at the cell surface cannot be detected easily. With citrate, nanoceria aggregates of size 20 to 500 nm are visible and interact with the cellular membrane and with microvili. Adsorbed particles are marked in yellow. Energy dispersive X-ray (EDX) analysis confirms that the particles contain cerium atoms.*

*Transmission Electron Microscopy (TEM):* Fibroblasts seeded with cerium oxides were further investigated by TEM. Fig. 6a and 6d provides representative images of NIH/3T3 cells incubated with $PAA_{2K}$–$CeO_2$ and to Cit–$CeO_2$, respectively. The experimental conditions were an incubation time of 24 h and a cerium concentration of 1 mM. A careful analysis of the TEM images shows that the particles were primarily located in membrane-bound compartments, or endosomes (Limbach et al. 2005b; Horie et al. 2011; Yokel et al. 2012). Close-up views of the selected areas (indicated by rectangles) identify the lipidic membrane separating the cytosol from the particles (Fig. 6b and 6e). In this work, nanoceria were found neither in the cytosol nor in the nucleus. A statistical analysis of the endosome sizes was performed and revealed notable variations as a function of the coating. For $PAA_{2K}$–$CeO_2$, the average size was 560 nm and similar to those of the control cell (data not shown), whereas for Cit–$CeO_2$, the endosomal distribution was peaked at 760 nm (Fig. 6c and 6f). With citrate, compartments larger than 1 µm were also detected. Another difference between the two coating lies in the spatial distribution of the particles inside endosomes: with citrate, the particles appear as aggregated under the form of clusters, whereas with polymers they are randomly spread and unassociated. With the $PAA_{2K}$-coated particles, the endosomes were also more homogeneously filled. In conclusion to this part, we have found that the carboxylate coated particles are internalized by the NIH/3T3 fibroblasts and located in endosomal membrane-bound compartments (Limbach et al. 2005b; Horie et al. 2011; Yokel et al. 2012; Hillaireau and Couvreur 2009; Safi et al. 2011). The similitudes between the compartments in Figs. 6a and 6d suggest similar mechanisms of entry into the cells, *i.e.* endocytosis (Iversen et al. 2011). The coating has here a definite but moderate impact on the endosome size distributions.

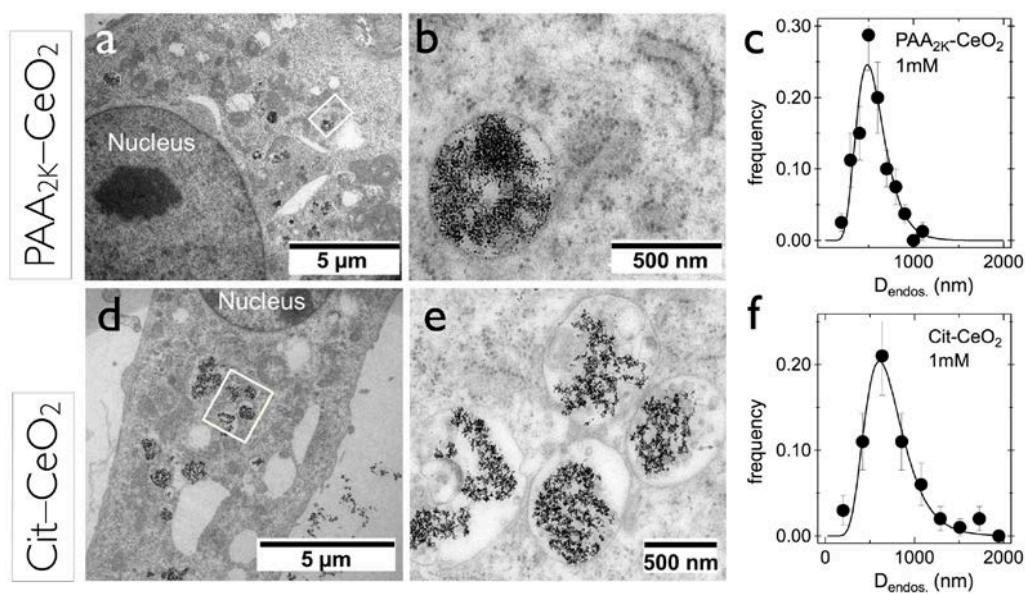

***Figure 6***: *Transmission electron microscopy images of NIH/3T3 fibroblast cells incubated with $PAA_{2K}$–$Ce_2O$ (a,b) and with Cit–$Ce_2O$ (d,e). The experimental conditions were an incubation time of 24 h and a cerium concentration 1 mM. The close views of the delimited areas in a) and d) show that the particles are localized in membrane-bound compartments of endosome type (b,e). The size distributions of the endosomes are shown in c) and f), respectively. With the units used in this work, [Ce] = 1 mM corresponds to c($Ce_2O$) = 0.017 wt. % or 172 µg mL$^{-1}$.*

## Flow Cytometry

*Cell viability:* The membrane integrity of cells treated with nanoparticles was assessed by flow cytometry after propidium iodide staining. Figs. 7a and 7b show the cell mortality after incubation times of 5 h and 24 h for cerium concentration 0.1, 1 and 10 mM. The cell mortality is compared to that of the organic molecules used to coat the particles, sodium citrate and poly(sodium acrylate). For the coating, the concentrations indicated in the figure are those of citrates and of acrylate monomers. They are representative of the amounts of citrates and $PAA_{2K}$ adsorbed onto nanoceria during the incubation with nanoparticles. After a short exposure, the cytotoxicity levels are low and comparable to that of the control. The highest value was obtained with $Cit\text{-}CeO_2$ at 10 mM, 2.7% *versus* 0.8% for the control. After 24 h, a noticeable increase of the mortality (21%) is observed for the citrate-coated particles at the highest dose. The results of Fig. 7 are in line with those testing the cellular growth (cell counting) and metabolic pathway (MTT) in the same conditions (Safi et al. 2010). MTT assays were recently performed on 2139 human lymphoblastoid cells to examine the impact of the cell line on the toxicity and the results were similar to those found for the fibroblasts (S8). The coating molecules exposed separately to the fibroblasts exhibit no noticeable toxicity (Fig. 7a).

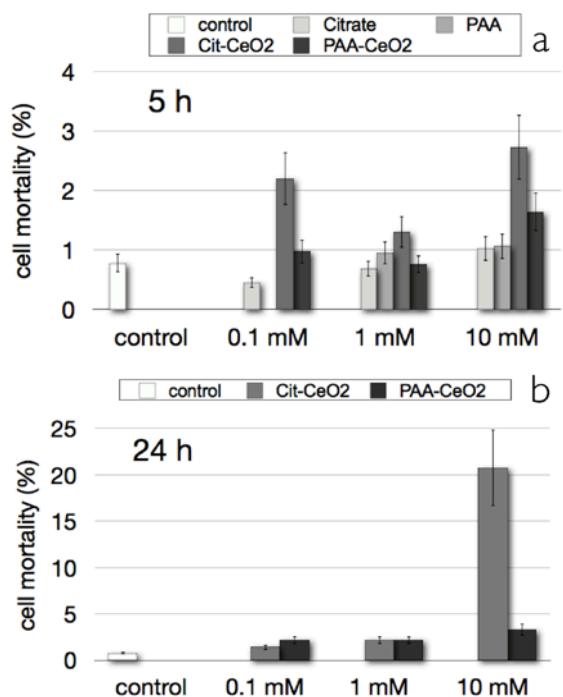

**Figure 7 :** *Cell mortality of NIH/3T3 fibroblasts as a function of the dose for incubation times 5 h (a) and 24 h (b) as determined from flow cytometry and propidium iodide staining. Data are compared to that of the organic molecules used to coat the particles, sodium citrate and poly(sodium acrylate) (a). Significant increase in mortality is found for $Cit\text{-}CeO_2$ at high dose (10 mM).*

*Side and Forward Scatter intensities from Flow Cytometry*: In flow cytometry experiments, the forward (FS) and side (SS) scatter intensities generated by the cell illumination are probing the cellular size and refractive index respectively. Figs. 8a and 8b display the SS intensities (median values) of NIH/3T3 that were incubated 5 and 24 h with nanoparticles at T = 37 °C. Living and dead cells were analyzed separately. Dead cells were distinguished by their strong red fluorescence linked to the nuclear incorporation of propidium iodide. The data are compared to those of untreated cells. Citrate-coated particles exhibit a shift of the SS signal to higher values with increasing dose.

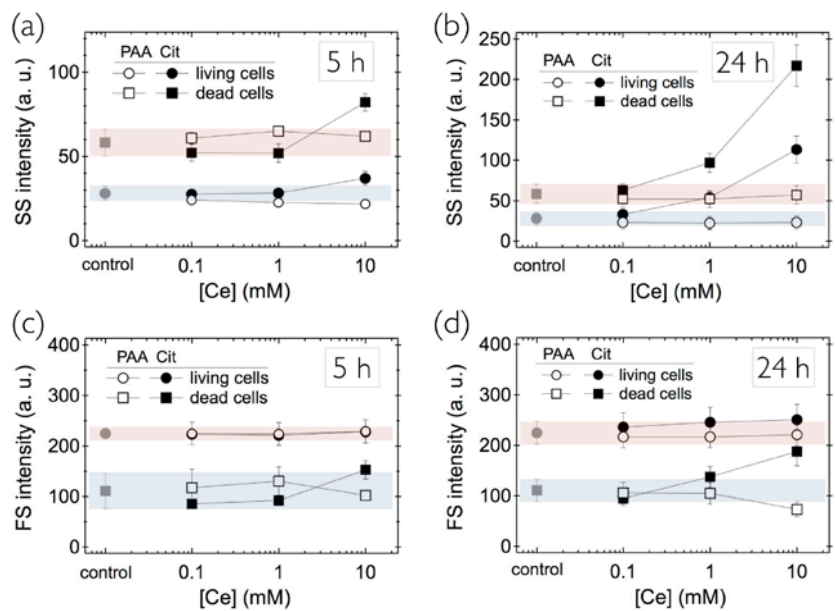

***Figure 8 :*** *Forward (a,b) and side (c,d) scatter intensities of NIH/3T3 incubated 5 and 24 h with nanoparticles at T = 37 °C measured by cytometry. Living and dead cells were analyzed separately (see Fig. 2 for details). Dead cells were distinguished by their strong red fluorescence linked to the nuclear incorporation of propidium iodide. Citrate-coated particles exhibit a shift of the SS signal to higher values with increasing dose, indicating strong nanoparticle/cell interactions. Cytometric intensities of cells seeded with the organic coating, either citrate or poly(sodium acrylate) display no variation as a function of the dose (S9). The red and blue bands in the 4 figures correspond to the values of the controls including the error bars.*

The shift remains moderate at 5 h, and becomes significant after a 24 h-incubation. The ratio noted $R_{SS}$ between the SS scatter of seeded cells and that of the control increases by a factor 3.4 for both living and dead cells. In Fig. 8b, the SS signal increases from 33 to 113 in arbitrary units and from 63 to 217, respectively. This result indicates that the refractive index of fibroblasts was enhanced thanks to their interactions with the particles. Comparable $R_{SS}$ ratios for dead and live cells at 5 and 24 h suggest that the nanoparticles interact in a similar fashion with both types of cells. These interactions are of two kinds, the adsorption on the plasma membrane (Fig. 5) and the internalization into endocytic vesicles (Fig. 6). With cytometry, equivalent results were reported on Chinese hamster ovary cells (Suzuki et al. 2007) treated with titanium dioxide, on human hepatoma (Xia et al. 2008a) and lymphablastoid (Safi et al. 2011) cells seeded with magnetic nanoparticles. By contrast, data for the polymer-coated particles show no evolution as a function of the cerium concentration ($R_{SS} \sim 1$). The absence of variations of the SS intensity for the $PAA_{2K}$–$CeO_2$ indicates low levels of nanomaterials interacting with the cells.

Figs. 8c and 8d provide FS intensities obtained in the same experimental conditions as those mentioned above. Within the experimental uncertainties, the FS signals for both living and dead cells remained unchanged at all incubation dose and time. These findings indicate that the size of the cells was not modified by the interactions with the particles regardless of the incubation time and dose (Suzuki et al. 2007; Xia et al. 2008a; Safi et al. 2011). For sake of completeness, it should be added that both SS and FS intensities of cells seeded with the organic coating, either citrate or poly(sodium acrylate) display no variation as a function of the dose (S9).

Reactive Oxygen Species: Reactive oxygen species (ROS) production was evaluated using the fluorescence changes resulting from oxidation of the permeant dye dihydroethidium (DHE) by intracellular superoxide anions. DHE exhibits blue-fluorescence in the cytosol until it is oxidized, while oxidized products intercalate within the DNA and exhibit bright red fluorescence. Cells were treated for 6 h or 24 h at different nanoparticle concentrations ([Ce] = 0.1, 1 and 10 mM). Negative and positive controls consisting of untreated cells or cells incubated for 2 and 4 hours with pyocyanin (Horke et al. 2010). Additional control experiments were performed after incubation with $PAA_{2K}$ polymers or citrate ligands. Individual cell red fluorescence and scatter properties were analyzed by flow cytometry. Quantification of the mean red fluorescence intensity of living cells (gated on double scatter histogram FS *versus* SS) as a function of the cerium oxide concentration revealed that exposure to nanomaterials modified only slightly the oxidation level of the fluorescent probe due to intracellular ROS activity (Fig. 9). Both untreated and cells exposed to nanomaterials generated a low fluorescence intensity (mean values ranging from 45 to 75 in arbitrary units) which were inferior to those of positive control (200 in arbitrary units). These results indicate that 6 h or 24 h treatment with coated particles did not significantly alter the production of reactive oxygen species by NIH/3T3 fibroblasts.

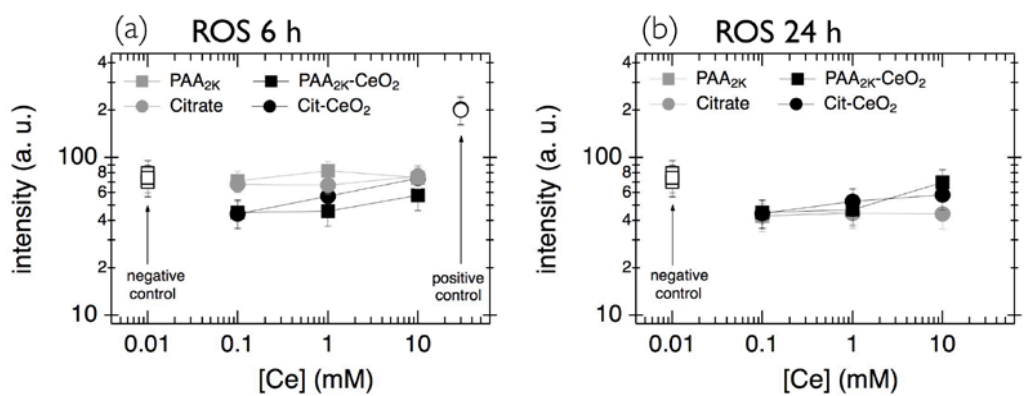

**Figure 9:** *Quantification of the mean red fluorescence intensity of living NIH/3T3 fibroblasts (gated on double scatter histogram FS versus SS, see Fig. 2) as a function of the cerium oxide concentration incubated for 6 h (a) and 24 h (b). Control experiments performed with the coating materials ($PAA_{2K}$ polymers and citrate ligands) are included for comparison. For ROS assays, the mean values of the red fluorescence intensities were used as a measure of the intracellular ROS activity.*

Comet Assays: DNA damage caused by nanoceria was studied using comet assay. Comet assays allow to reveal breakage of single or double strands DNA and chromosome rearrangement. The fibroblasts were treated with the Cit-CeO₂ particles during 4 h at 1 and 10 mM, and compared to negative (untreated cells) and positive (pyocyanin) controls (Fig. 10). As illustrated in Fig. 10b, a comet-like tail implies the presence of DNA strands that were dragged out of the nucleus by the electrophoretic field. The extension of the tail increases with the degree of damage. Tail momentum of control DNA was compared with nanoparticle treated cells, and extent of damage was assessed in Fig. 10d. For treated cells, the comet tails remained in the lowest range as compared to the negative control. A slight but significant increase in DNA damage was observed between the 1 and 10 mM assays however.

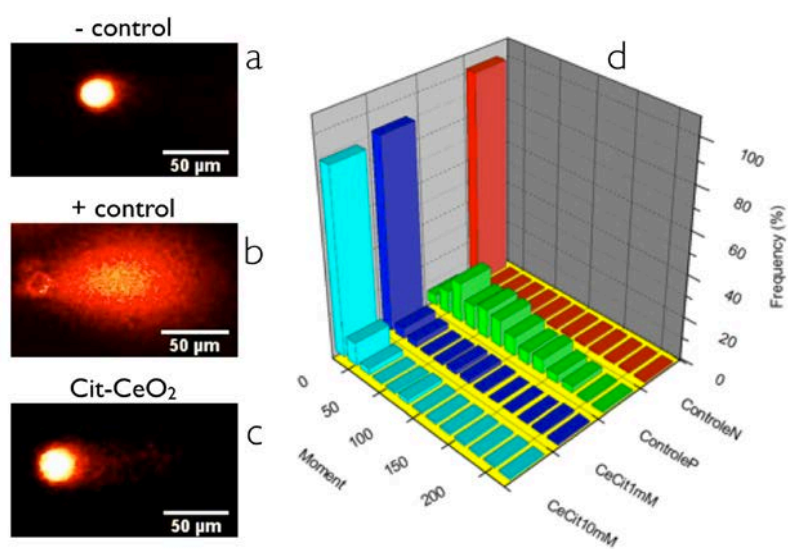

**Figure 10:** *Comet assays carried out on NIH/3T3 fibroblasts treated with the Cit-CeO₂ particles during 4 h at [Ce] = 1 and 10 mM. Negative (untreated cells) and positive (H₂O₂) controls are shown for comparison. a), b) and c) Illustrations of tail momentum of DNA strands that were dragged out of the nucleus by the electrophoretic field for negative and positive controls and for the particles. d) Analysis of the tail momentum for the different samples.*

# Discussion

In this work, the interactions and the short-term toxicity of nanoceria with respect to NIH/3T3 fibroblasts were investigated combining light scattering, cytometry and electron microscopy. The cytotoxicity of citrate and poly(acrylic acid) acid coated particles was examined. In a first report (Safi et al. 2010), the potential adverse effects of these particles were examined using cell proliferation and MTT assays. Both methods revealed a dose-dependent toxicity for the citrate coated particles (above 3 mM or 500 µg mL⁻¹). Here, we examined these particles in a broader toxicological context, and focus on their interactions at the scale of the cell. At each step of the interaction process, the particles were monitored and localized thanks to various techniques, including optical, scanning and transmission electron microscopy. Nanoceria coated with citrates were located at the plasma membrane in the form of single particles or of 20 – 500 nm aggregates. Inside cells, both particle types were found in membrane-bound compartments. These findings are in agreement with several reports from the literature (Horie et al. 2011; Yokel et al. 2012). In terms of toxicity, the membrane integrity of treated fibroblasts was evaluated using propidium iodide staining and flow cytometry. Cytometry confirmed previous results on the cellular growth and metabolic pathway: coated nanoceria do not display short-term acute toxicity. At 24 h, a noticeable increase of the mortality (21%) was however observed for the citrate-coated particles, but only at the highest dose (10 mM). With $PAA_{2K}$-$CeO_2$ the cell viability remained high at all doses. In the presence of the permeant dye dihydroethidium, treated particles did not exhibit significant fluorescence, indicating a low level of induced oxidative stress. Additional testing using non-adherent human lymphoblasts were conducted to probe the cell line dependence, and comparable results were observed. Although the nanoceria concentrations investigated here are of the same order as those found in the literature, it should be mentioned that the doses are far above realistic values, especially if one considers the time scale of the toxicity assays. In other terms, there is a gap between the experimental conditions to which the cells were exposed and real world exposures. On a longer term, continual exposure could lead to an increased load of nanomaterials in cells and tissues and concentrations may indeed get closer to levels studied here.

The case of citrate coated particles deserves further comments. First, the present results confirm conclusions already formulated in the literature, namely that low-molecular weight ligands with reversible adsorption properties, such as citrate ions are poor coating agents for particles in biological environments (Kuckelhaus et al. 2003; Ojea-Jimenez and Puntes 2009; Maiorano et al. 2010; Safi et al. 2011; Galimard et al. 2012). As shown in Fig. 3, the particles precipitate instantaneously in cell media. In culture conditions, the gravitational settling of the precipitating particles enhances the bioavailability of the cell, and one may ask whether the interaction with the cells is not simply driven by this sedimentation process. Cytometry and SEM (data not shown) performed as a function of the incubation time, as well as data obtained on similarly coated particles (Safi et al. 2011; Galimard et al. 2012) support a scenario in which nanoceria brought in contact with the biofluid also precipitate on the cell membranes. In the present experiments, sedimentation of precipitated particles occurs within the first hours of incubation, typically from 2 to 12 hours. However, citrate coated nanoceria were shown to adsorb and be internalized after few minutes of incubation. A possible interpretation is that the aggregates that form in biofluids are neutral or positively charged, and thus interact more strongly with the cell membrane. On a longer time scale, the sedimentation enters into play and further increases the adsorption (Limbach et al. 2005a; Teeguarden et al. 2007).

As mentioned in the introduction, reports have been published describing controversial results on the toxicity of cerium oxide. *In vitro* and *in vivo* assays have demonstrated in some cases highly toxic effects of nanoceria, even at very low concentrations (< 1µM) (Eom and Choi 2009; Zhang et al. 2011)), and in others a relatively benign impact on living environments (Xia et al. 2008b; Park et al. 2010; Safi et al. 2010; Birbaum et al. 2010; Horie et al. 2012; Pierscionek et al. 2012). Comparing a wide variety of data collected recently on nanoceria, Karakoti *et al.* suggested that the thermal history of the dispersion (*i.e.* during synthesis) could have an influence on their biological impact (Karakoti et al. 2010). According to these authors, the thermal processing may alter many physicochemical properties of the particles, and explain why $CeO_2$ can have pro- or anti-oxidative behaviors, or be neutral with respect to cells. In this work, we have found that a second factor, namely the coating and its ability to protect from aggregation can affect the toxicity of these particles. The cytometry data in Fig. 8 demonstrated that citrate-coated particles interact more strongly with the fibroblasts, as the side-scatter intensity was increased by the uptake of the particles. We anticipate that for citrate-coated particles, the enhanced interactions resulted in a net toxicity. These results demonstrate that a control of the surface chemistry of the particles is crucial when dealing with toxicity and nanoparticle/cell interactions.

# Acknowledgments

We thank Armelle Baeza-Squiban, Jean-Paul Chapel, Gaelle Charron, Emek Seyrek, Sonja Boland for fruitful discussions. We thank Frédéric Herbst from the Laboratoire Interdisciplinaire Carnot de Bourgogne for his help in the scanning electron microscopy studies and Virginie Garnier-Thibaud for the TEM measurements at the microscopic platform of the University Pierre et Marie Curie (Paris 6). ANR (Agence Nationale de la Recherche) and CGI (Commissariat à l'Investissement d'Avenir) are gratefully acknowledged for their financial support of this work through Labex SEAM (Science and Engineering for Advanced Materials and devices). This research was supported in part by the Agence Nationale de la Recherche under the contract ANR-09-NANO-P200-36.

# Supporting Information

The Supporting Information section provides complementary data on the structure of the $CeO_2$ nanoparticles (S1), the electrophoretic mobility measurements of the particles of different coating (S2), on the absorptivity coefficient of cerium oxide (S3), the stability of the Cit-$CeO_2$ particles in PBS at short and long time (S4), the light scattering properties of the cell culture medium used in this study (DMEM) (S5), the diameter and length distributions of microvilli at the membrane of NIH/3T3 fibroblasts (S6), the energy Dispersive X-Ray Spectrometry and Scanning Electron Microscopy (S7), MTT assays performed on NIH/3T3 mouse fibroblasts and 2139 human lymphoblastoid cells (S8) and the finally side and forward scatter intensities obtained by flow cytometry (S9). This information is available free of charge *via* the Internet at xxx.

# References

Auffan M, Rose J, Orsiere T, De Meo M, Thill A, Zeyons O, Proux O, Masion A, Chaurand P, Spalla O, Botta A, Wiesner MR, Bottero JY. 2009. CeO2 nanoparticles induce DNA damage towards human dermal fibroblasts in vitro. Nanotoxicology 3:161-171.

Berret J-F. 2007. Stoichiometry of electrostatic complexes determined by light scattering. Macromolecules 40:4260-4266.

Berret J-F, Sandre O, Mauger A. 2007. Size distribution of superparamagnetic particles determined by magnetic sedimentation. Langmuir 23:2993-2999.

Birbaum K, Brogioli R, Schellenberg M, Martinoia E, Stark WJ, Günther D, Limbach LK. 2010. No Evidence for Cerium Dioxide Nanoparticle Translocation in Maize Plants. Environmental Science & Technology 44:8718-8723.

Brunner TJ, Wick P, Manser P, Spohn P, Grass RN, Limbach LK, Bruinink A, Stark WJ. 2006. In Vitro Cytotoxicity of Oxide Nanoparticles: Comparison to Asbestos, Silica, and the Effect of Particle Solubility. Environ. Sci. Technol. 40:4374 - 4381.

Casals E, Pfaller T, Duschl A, Oostingh GJ, Puntes V. 2010. Time Evolution of the Nanoparticle Protein Corona. Acs Nano 4:3623-3632.

Cassee FR, van Balen EC, Singh C, Green D, Muijser H, Weinstein J, Dreher K. 2011. Exposure, Health and Ecological Effects Review of Engineered Nanoscale Cerium and Cerium Oxide Associated with its Use as a Fuel Additive. Critical Reviews in Toxicology 41:213-229.

Chanteau B, Fresnais J, Berret J-F. 2009. Electrosteric Enhanced Stability of Functional Sub-10 nm Cerium and Iron Oxide Particles in Cell Culture Medium. Langmuir 25:9064-9070.

Chigurupati S, Mughal MR, Okun E, Das S, Kumar A, McCaffery M, Seal S, Mattson MP. 2013. Effects of cerium oxide nanoparticles on the growth of keratinocytes, fibroblasts and vascular endothelial cells in cutaneous wound healing. Biomaterials 34:2194-2201.

Cho WS, Duffin R, Poland CA, Howie SEM, MacNee W, Bradley M, Megson IL, Donaldson K. 2010. Metal Oxide Nanoparticles Induce Unique Inflammatory Footprints in the Lung: Important Implications for Nanoparticle Testing. Environmental Health Perspectives 118:1699-1706.

Culcasi M, Benameur L, Mercier A, Lucchesi C, Rahmouni H, Asteian A, Casano G, Botta A, Kovacic H, Pietri S. 2012. EPR spin trapping evaluation of ROS production in human fibroblasts exposed to cerium oxide nanoparticles: Evidence for NADPH oxidase and mitochondrial stimulation. Chem.-Biol. Interact. 199:161-176.

Das M, Patil S, Bhargava N, Kang J-F, Riedel LM, Seal S, Hickman JJ. 2007. Auto-catalytic ceria nanoparticles offer neuroprotection to adult rat spinal cord neurons. Biomaterials 28:1918 - 1925.

Diaz B, Sanchez-Espinel C, Arruebo M, Faro J, de Miguel E, Magadan S, Yague C, Fernandez-Pacheco R, Ibarra MR, Santamaria J, Gonzalez-Fernandez A. 2008. Assessing Methods for Blood Cell Cytotoxic Responses to Inorganic Nanoparticles and Nanoparticle Aggregates. Small 4:2025-2034.

Eom HJ, Choi J. 2009. Oxidative stress of CeO2 nanoparticles via p38-Nrf-2 signaling pathway in human bronchial epithelial cell, Beas-2B. Toxicology Letters 187:77-83.

Eugene E, Hoffmann I, Pujol C, Couraud PO, Bourdoulous S, Nassif X. 2002. Microvilli-like structures are associated with the internalization of virulent capsulated Neisseria meningitidis into vascular endothelial cells. Journal of Cell Science 115:1231-1241.

Galimard A, Safi M, Ould-Moussa N, Montero D, Conjeaud H, Berret JF. 2012. Thirty-Femtogram Detection of Iron in Mammalian Cells. Small 8:2036-2044.

Hecht E, Usmani SM, Albrecht S, Wittekindt OH, Dietl P, Mizaikoff B, Kranz C. 2011. Atomic force microscopy of microvillous cell surface dynamics at fixed and living alveolar type II cells. Analytical and Bioanalytical Chemistry 399:2369-2378.

HEI. 2001. Evaluation of Human Health Risk from Cerium Added to Diesel Fuel. In: Comm. 9 p-, ed. Comm. 9, pp 1-57. Health Effects Institute (HEI).

Hillaireau H, Couvreur P. 2009. Nanocarriers' Entry into the Cell: Relevance to Drug Delivery. Cellular and Molecular Life Sciences 66:2873-2896.

Hirst SM, Karakoti AS, Tyler RD, Sriranganathan N, Seal S, Reilly CM. 2009. Anti-inflammatory Properties of Cerium Oxide Nanoparticles. Small 5:2848-2856.

Horie M, Kato H, Fujita K, Endoh S, Iwahashi H. 2012. In Vitro Evaluation of Cellular Response Induced by Manufactured Nanoparticles. Chemical Research in Toxicology 25:605-619.

Horie M, Nishio K, Kato H, Fujita K, Endoh S, Nakamura A, Miyauchi A, Kinugasa S, Yamamoto K, Niki E, Yoshida Y, Hagihara Y, Iwahashi H. 2011. Cellular responses induced by cerium oxide nanoparticles: induction of intracellular calcium level and oxidative stress on culture cells. Journal of Biochemistry 150:461-471.

Horke S, Witte I, Altenhofer S, Wilgenbus P, Goldeck M, Forstermann U, Xiao JH, Kramer GL, Haines DC, Chowdhary PK, Haley RW, Teiber JF. 2010. Paraoxonase 2 is down-regulated by the Pseudomonas aeruginosa quorum-sensing signal N-(3-oxododecanoyl)-L-homoserine lactone and attenuates oxidative stress induced by pyocyanin. Biochemical Journal 426:73-83.

Hussain S, Al-Nsour F, Rice AB, Marshburn J, Ji ZX, Zink JI, Yingling B, Walker NJ, Garantziotis S. 2012a. Cerium dioxide nanoparticles do not modulate the lipopolysaccharide-induced inflammatory response in human monocytes. International Journal of Nanomedicine 7:1387 - 1397.

Hussain S, Al-Nsour F, Rice AB, Marshburn J, Yingling B, Ji Z, Zink JI, Walker NJ, Garantziotis S. 2012b. Cerium Dioxide Nanoparticles Induce Apoptosis and Autophagy in Human Peripheral Blood Monocytes. ACS Nano 6:5820 - 5829.

Iversen TG, Skotland T, Sandvig K. 2011. Endocytosis and intracellular transport of nanoparticles: Present knowledge and need for future studies. Nano Today 6:176-185.

Karakoti A, Singh S, Dowding JM, Seal S, Self WT. 2010. Redox-active radical scavenging nanomaterials. Chemical Society Reviews 39:4422-4432.

Karakoti AS, Munusamy P, Hostetler K, Kodali V, Kuchibhatla S, Orr G, Pounds JG, Teeguarden JG, Thrall BD, Baer DR. 2012. Preparation and characterization challenges to understanding environmental and biological impacts of ceria nanoparticles. Surface and Interface Analysis 44:882-889.

Kemp SJ, Thorley AJ, Gorelik J, Seckl MJ, O'Hare MJ, Arcaro A, Korchev Y, Goldstraw P, Tetley TD. 2008. Immortalization of Human Alveolar Epithelial Cells to Investigate Nanoparticle Uptake. American Journal of Respiratory Cell and Molecular Biology 39:591-597.

Koeneman BA, Zhang Y, Westerhoff P, Chen YS, Crittenden JC, Capco DG. 2010. Toxicity and cellular responses of intestinal cells exposed to titanium dioxide. Cell Biology and Toxicology 26:225-238.

Kroll A, Dierker C, Rommel C, Hahn D, Wohlleben W, Schulze-Isfort C, Gobbert C, Voetz M, Hardinghaus F, Schnekenburger J. 2011. Cytotoxicity screening of 23 engineered nanomaterials using a test matrix of ten cell lines and three different assays. Part. Fibre Toxicol. 8.

Kuckelhaus S, Garcia VAP, Lacava LM, Azevedo RB, Lacava ZGM, Lima ECD, Figueiredo F, Tedesco AC, Morais PC. 2003. Biological investigation of a citrate-coated cobalt-ferrite-based magnetic fluid. Journal of Applied Physics 93:6707-6708.

Lee SS, Zhu HG, Contreras EQ, Prakash A, Puppala HL, Colvin VL. 2012. High Temperature Decomposition of Cerium Precursors To Form Ceria Nanocrystal Libraries for Biological Applications. Chem. Mat. 24:424-432.

Limbach LK, Li Y, Grass RN, Brunner TJ, Hintermann MA, Muller M, Gunther D, Stark WJ. 2005a. Oxide Nanoparticle Uptake in Human Lung Fibroblasts: Effects of Particle Size, Agglomeration, and Diffusion at Low Concentrations. Environmental Science & Technology 39:9370-9376.

Limbach LK, Li Y, Grass RN, Brunner TJ, Hintermann MA, Muller M, Gunther D, Stark WJ. 2005b. Oxide Nanoparticle Uptake in Human Lung Fibroblasts: Effects of Particle Size, Agglomeration, and Diffusion at Low Concentrations. Environmental Science & Technology 39:9370 - 9376.

Lin WS, Huang YW, Zhou XD, Ma YF. 2006. Toxicity of cerium oxide nanoparticles in human lung cancer cells. International Journal of Toxicology 25:451-457.

Lorenz MR, Holzapfel V, Musyanovych A, Nothelfer K, Walther P, Frank H, Landfester K, Schrezenmeier H, Mailander V. 2006. Uptake of functionalized, fluorescent-labeled polymeric particles in different cell lines and stem cells. Biomaterials 27:2820-2828.

Ma JY, Mercer RR, Barger M, Schwegler-Berry D, Scabilloni J, Ma JK, Castranova V. 2012. Induction of pulmonary fibrosis by cerium oxide nanoparticles. Toxicology and Applied Pharmacology 262:255-264.

Maiorano G, Sabella S, Sorce B, Brunetti V, Malvindi MA, Cingolani R, Pompa PP. 2010. Effects of Cell Culture Media on the Dynamic Formation of Protein-Nanoparticle Complexes and Influence on the Cellular Response. Acs Nano 4:7481-7491.

Nabavi M, Spalla O, Cabane B. 1993. Surface Chemistry of Nanometric Ceria Particles in Aqueous Dispersions. J. Colloid Interface Sci. 160:459 - 471.

Ojea-Jimenez I, Puntes V. 2009. Instability of Cationic Gold Nanoparticle Bioconjugates: The Role of Citrate Ions. Journal of the American Chemical Society 131:13320-13327.

Park B, Donaldson K, Duffin R, Tran L, Kelly F, Mudway I, Morin JP, Guest R, Jenkinson P, Samaras Z, Giannouli M, Kouridis H, Martin P. 2008. Hazard and risk assessment of a nanoparticulate cerium oxide-based diesel fuel additive - A case study. Inhalation Toxicology 20:547-566.

Park EJ, Cho WS, Jeong J, Yi JH, Choi K, Kim Y, Park K. 2010. Induction of Inflammatory Responses in Mice Treated with Cerium Oxide Nanoparticles by Intratracheal Instillation. Journal of Health Science 56:387-396.

Petri-Fink A, Steitz B, Finka A, Salaklang J, Hofmann H. 2008. Effect of cell media on polymer coated superparamagnetic iron oxide nanoparticles (SPIONs): Colloidal stability, cytotoxicity, and cellular uptake studies. European Journal of Pharmaceutics and Biopharmaceutics 68:129-137.

Pierscionek BK, Li YB, Schachar RA, Chen W. 2012. The effect of high concentration and exposure duration of nanoceria on human lens epithelial cells. Nanomedicine-Nanotechnology Biology and Medicine 8:383-390.

Qi L, Chapel JP, Castaing JC, Fresnais J, Berret J-F. 2008a. Organic versus hybrid coacervate complexes: co-assembly and adsorption properties. Soft Matter 4:577-585.

Qi L, Sehgal A, Castaing JC, Chapel JP, Fresnais J, Berret J-F, Cousin F. 2008b. Redispersible hybrid nanopowders: Cerium oxide nanoparticle complexes with phosphonated-PEG oligomers. Acs Nano 2:879-888.

Rocker C, Potzl M, Zhang F, Parak WJ, Nienhaus GU. 2009. A quantitative fluorescence study of protein monolayer formation on colloidal nanoparticles. Nature Nanotechnology 4:577-580.

Safi M, Courtois J, Seigneuret M, Conjeaud H, Berret J-F. 2011. The effects of aggregation and protein corona on the cellular internalization of iron oxide nanoparticles. Biomaterials 32:9353-9363.

Safi M, Sarrouj H, Sandre O, Mignet N, Berret J-F. 2010. Interactions between sub-10-nm iron and cerium oxide nanoparticles and 3T3 fibroblasts: the role of the coating and aggregation state. Nanotechnology 21:145103.

Simonelli F, Marmorato P, Abbas K, Ponti J, Kozempel J, Holzwarth U, Franchini F, Rossi F. 2011. Cyclotron Production of Radioactive CeO2 Nanoparticles and Their Application for In Vitro Uptake Studies. IEEE Trans. Nanobiosci. 10:44-50.

Srinivas A, Rao PJ, Selvam G, Murthy PB, Reddy PN. 2011. Acute inhalation toxicity of cerium oxide nanoparticles in rats. Toxicology Letters 205:105-115.

Sund J, Alenius H, Vippola M, Savolainen K, Puustinen A. 2011. Proteomic Characterization of Engineered Nanomaterial-Protein Interactions in Relation to Surface Reactivity. Acs Nano 5:4300-4309.

Suzuki H, Toyooka T, Ibuki Y. 2007. Simple and easy method to evaluate uptake potential of nanoparticles in mammalian cells using a flow cytometric light scatter analysis. Environmental Science & Technology 41:3018-3024.

Teeguarden JG, Hinderliter PM, Orr G, Thrall BD, Pounds JG. 2007. Particokinetics in vitro: Dosimetry considerations for in vitro nanoparticle toxicology assessments. Toxicological Sciences 97:614-614.

Thanh NTK, Green LAW. 2010. Functionalisation of nanoparticles for biomedical applications. Nano Today 5:213-230.

Thill A, Zeyons O, Spalla O, Chauvat F, Rose J, Auffan M, Flank AM. 2006. Cytotoxicity of CeO2 nanoparticles for Escherichia coli. Physico-chemical insight of the cytotoxicity mechanism. Environmental Science & Technology 40:6151-6156.

Vincent A, Babu S, Heckert E, Dowding J, Hirst SM, Inerbaev TM, Self WT, Reilly CM, Masunov AE, Rahman TS, Seal S. 2009. Protonated Nanoparticle Surface Governing Ligand Tethering and Cellular Targeting. Acs Nano 3:1203-1211.

Walczyk D, Bombelli FB, Monopoli MP, Lynch I, Dawson KA. 2010. What the Cell "Sees" in Bionanoscience. Journal of the American Chemical Society 132:5761-5768.

Walser T, Limbach LK, Brogioli R, Erismann E, Flamigni L, Hattendorf B, Juchli M, Krumeich F, Ludwig C, Prikopsky K, Rossier M, Saner D, Sigg A, Hellweg S, Gunther D, Stark WJ. 2012. Persistence of engineered nanoparticles in a municipal solid-waste incineration plant. Nature Nanotechnology 7:520-524.

Williams D, Ehrman S, Pulliam Holoman T. 2006. Evaluation of the microbial growth response to inorganic nanoparticles. Journal of Nanobiotechnology 4:3 - 10.

Xia J, Zhang S, Zhang Y, Ma M, Xu K, Tang M, Gu N. 2008a. The Relationship Between Internalization of Magnetic Nanoparticles and Changes of Cellular Optical Scatter Signal. Journal of Nanoscience and Nanotechnology 8:6310-6315.

Xia T, Kovochich M, Liong M, Madler L, Gilbert B, Shi HB, Yeh JI, Zink JI, Nel AE. 2008b. Comparison of the Mechanism of Toxicity of Zinc Oxide and Cerium Oxide Nanoparticles Based on Dissolution and Oxidative Stress Properties. Acs Nano 2:2121-2134.

Yokel RA, Au TC, MacPhail R, Hardas SS, Butterfield DA, Sultana R, Goodman M, Tseng MT, Dan M, Haghnazar H, Unrine JM, Graham UM, Wu P, Grulke EA. 2012. Distribution, Elimination, and Biopersistence to 90 Days of a Systemically Introduced 30 nm Ceria-Engineered Nanomaterial in Rats. Toxicological Sciences 127:256-268.

Yokel RA, Florence RL, Unrine JM, Tseng MT, Graham UM, Wu P, Grulke EA, Sultana R, Hardas SS, Butterfield DA. 2009. Biodistribution and oxidative stress effects of a systemically-introduced commercial ceria engineered nanomaterial. Nanotoxicology 3:234-248.

Zhang HF, He XA, Zhang ZY, Zhang P, Li YY, Ma YH, Kuang YS, Zhao YL, Chai ZF. 2011. Nano-CeO2 Exhibits Adverse Effects at Environmental Relevant Concentrations. Environmental Science & Technology 45:3725-3730.

Zhang W, Kalive M, Capco DG, Chen YS. 2010. Adsorption of hematite nanoparticles onto Caco-2 cells and the cellular impairments: effect of particle size. Nanotechnology 21:355103.

# Supporting Information

## *In vitro toxicity of nanoceria : effect of coating and stability in biofluids*

N. Ould-Moussa, M. Safi, M.-A. Guedeau-Boudeville, D. Montero, H. Conjeaud and J.-F. Berret[*]

---

*Outline*

---

**S1 – Structural properties of CeO₂ nanoparticles**
**S2 – Electrophoretic mobility measurements of coated CeO₂**
**S3 – Absorption coefficient of cerium oxide nanoparticles**
**S4 – Long time stability of the Cit-CeO₂ particles in PBS**
**S5 – Light scattering properties of cell culture medium**
**S6 – Diameter and length distributions of microvilli at the cell membrane**
**S7 – Energy Dispersive X-Ray Spectrometry and Scanning Electron Microscopy**
**S8 – MTT assays performed on NIH/3T3 mouse fibroblasts and 2139 human lymphoblastoid cells**
**S9 – Side and Forward Scatter Intensities as received from Flow Cytometry**

---

## S1 – Structural properties of CeO₂ nanoparticles

Wide-angle x-ray scattering on concentrated CeO₂ dispersion in acidic conditions was performed at the Laboratory Itodys (University Paris7 – Denis Diderot, Paris, France). The intensity exhibited up to 9 structural Bragg peaks. The sequence is in accordance with a face-centered cubic structure of CeO2. The lattice parameter of the crystal is 5.47 Å. Using the Debye-Scherrer relationship that relates the full width at half maximum to the spatial extension of a crystalline microstructure, one obtained a value for the crystalline core of 5 nm. This value is in excellent agreement with the TEM results (Fig. 1 in main text). In Fig. S1, the positions of the Bragg reflections of the CeO₂ lattice are shown for comparison.

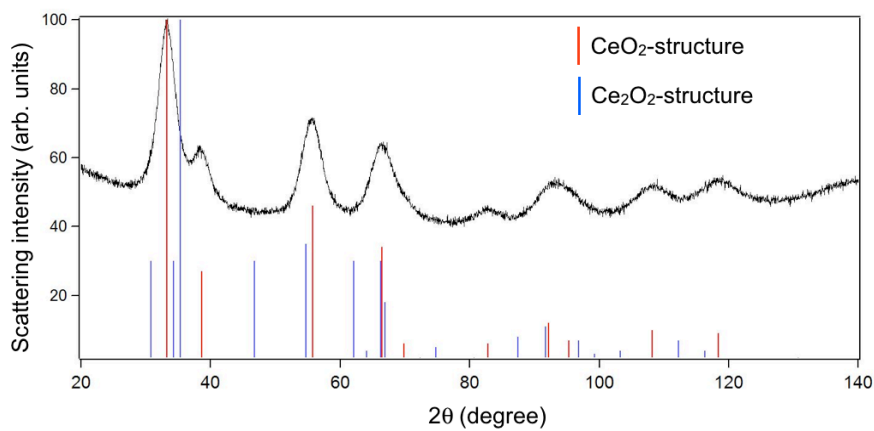

**Fig. S1**: *Crystalline structure of CeO₂ nanoparticles as studied by WAXS. The positions of the Bragg peaks are in good agreement with the predicted CeO₂-fluorite structure.*

## S2 – Electrophoretic mobility measurements of coated $CeO_2$

For the anionically charged particles, measurements of the zeta potential $\zeta$ and electrophoretic mobility $\mu_E$ were performed using a Zetasizer Nano ZS (Malvern Instrument). Using laser Doppler velocimetry, the technique is based on the Phase Analysis Light Scattering (PALS) method. The data for the two quantities are listed in Table S2 and show that the citrate and PAA-coated particles were negatively charged.

| nanoparticles | pH | Zeta potential $\zeta$ mV | Electrophoretic Mobility $\mu_E$ $10^{-4}$ cm$^2$ V$^{-1}$ |
|---|---|---|---|
| $CeO_2$ (uncoated) | 1.5 | + 40 | + 3.1 |
| Cit-$CeO_2$ | 8 | - 24 | - 1.87 |
| PAA$_{2K}$-$CeO_2$ | 8 | - 45 | - 3.52 |

**Table S2**: *Zeta potential ($\zeta$) and electrophoretic mobility ($\mu_E$) for uncoated and for anionically coated cerium oxide nanoparticles.*

## S3 – Absorption coefficient of cerium oxide nanoparticles

In Fig. S3 is shown the molar absorption coefficient $\varepsilon$ (cm$^{-1}$ M$^{-1}$) for cerium oxide nanoparticles. Using this calibration curve, the cerium concentration of dispersion was determined accurately. Citrate and polymer coating did not modify the absorption coefficient.

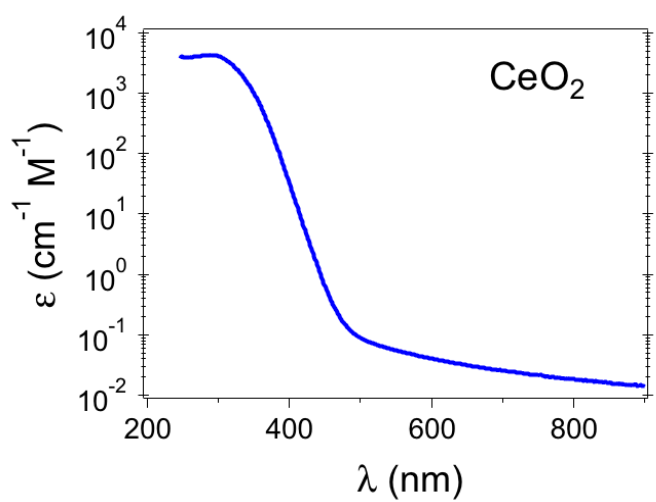

**Figure S3**: *Molar absorption coefficient of bare nanoceria.*

## S4 – Long time stability of the Cit-CeO₂ particles in PBS

In Fig. 3 of the main text, it was seen that the hydrodynamic diameter of citrate-coated nanoceria dispersed in PBS was slightly increasing with time. In two hours, $D_H$ was changing from 7 to 11 nm. To substantiate that this increase was the sign of a long-term destabilization, the light scattering experiment was repeated on a longer time period. Fig. S4 displays the $D_H$-variation over more than 8 hours. The regular increase demonstrates the slow destabilization of the particles in PBS. The blue area in the figure attests of the range where particle clusters start to sediment. Note that on the same time period poly(acrylic acid) coated nanoceria did not destabilize in the same buffer.

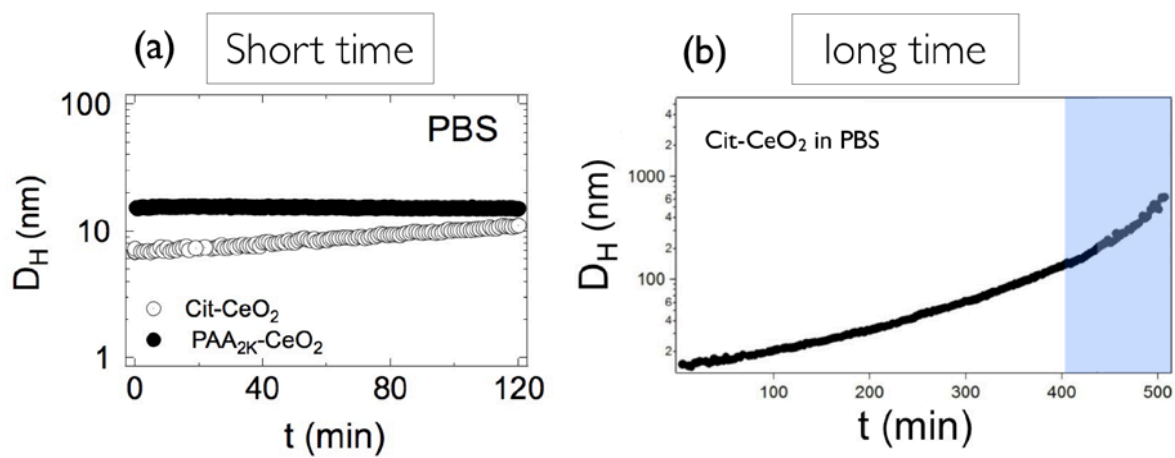

**Figure S4**: *Time dependence of the hydrodynamic diameter of citrate coated nanoceria at short and long times. The increase in $D_H$ is interpreted as the aggregation of the single particles into large sized clusters. The grey area on the right-hand diagram denotes the range where the aggregates started to sediment at the bottom of the light scattering tube.*

## S5 – Light scattering properties of cell culture medium

To evaluate the contribution arising form the proteins and other biological molecules dispersed in cellular media, we carried out a complete characterization of the cell media by dynamical light scattering. Fig. S5a displays the first order auto-correlation function of the scattered light for a 4 wt. % BSA (bovine serum albumin) sample, for DMEM with 10 vol. % fetal bovine serum (FBS) and for the medium of a cell culture. This later medium was obtained after centrifugation of lymphoblastoid cells (1200 rpm for 5 mn) and by pipetting the supernatant. The $g^{(1)}(t)$ functions for the two last specimen were similar, and with a relaxation time larger than that of the BSA proteins. A CONTIN analysis of the time traces yielded the size distributions shown in Fig. S5b-d. The intensity distribution was found to be peaked around 7 nm for the BSA, with an additional contribution at 30 nm, suggesting a possible aggregation of the BSA. By contrast, the cellular media exhibited broad size distributions centered near 10 nm, but extending to diameters as high as 100 nm.

Light scattering on cellular media

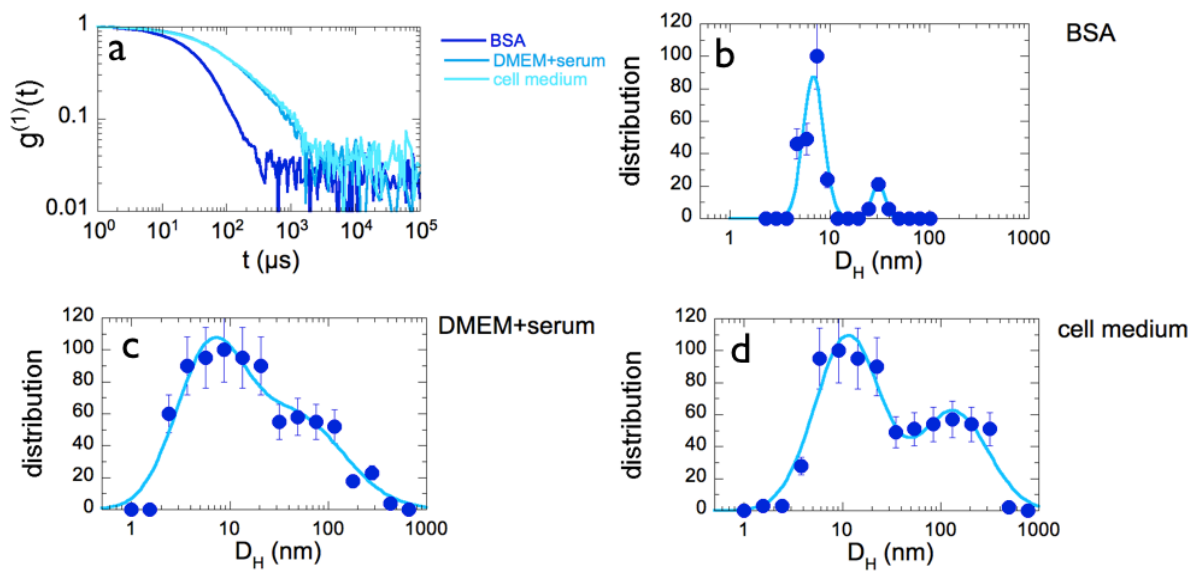

**Figure S5**: *a) First order auto-correlation function of the scattered light in double logarithmic scales, for a 4 wt. % BSA (bovine serum albumin) sample, for DMEM with 10 vol.% fetal bovine serum (FBS) and from a cell culture medium. b-d) Size distributions (based on the analysis of the scattering signal in terms of intensity) found for the three above solvents.*

## S6 – Diameter and length distributions of microvilli at the cell membrane

The diameter and length of the microvilli present at the surface of NIH/3T3 cells were measured on a sample of more than 100 protrusions. The distributions are displayed in Figs. S6a and S6b, respectively. Median diameter and length are 0.157 µm and 0.690 µm. The polydispersity of the length distribution is much larger than that of the diameter. Similar values were found for cells treated with Cit-$CeO_2$ and with $PAA_{2K}$-$CeO_2$.

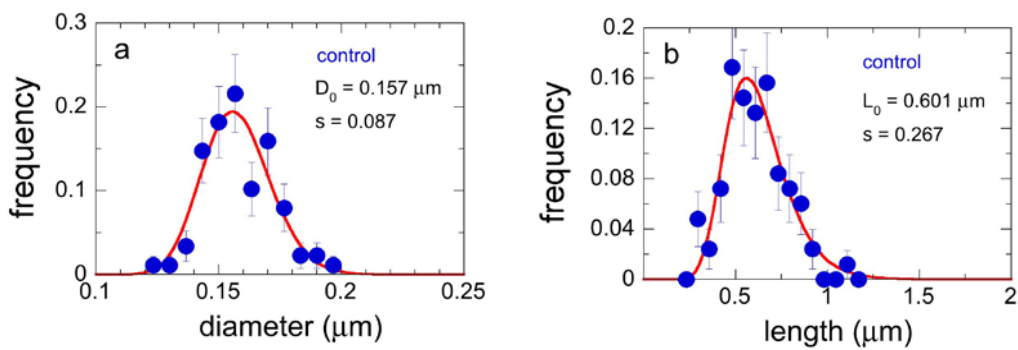

**Figure S6a and S6b**: *Distributions on the diameter and length of protrusions (microvilli) found at the surface of NIH/3T3 cells in control experiments. The incubation with nanoceria did not modify these distributions.*

## S7 – Energy Dispersive X-Ray Spectrometry and Scanning Electron Microscopy

Cerium was found at the surfaces of cells treated with both citrate and PAA$_{2K}$ coated particles. Due to the low amplitude of the cerium peak for the polymer-coated particles, quantitative modeling of the spectra was necessary to determine the cerium percentage.

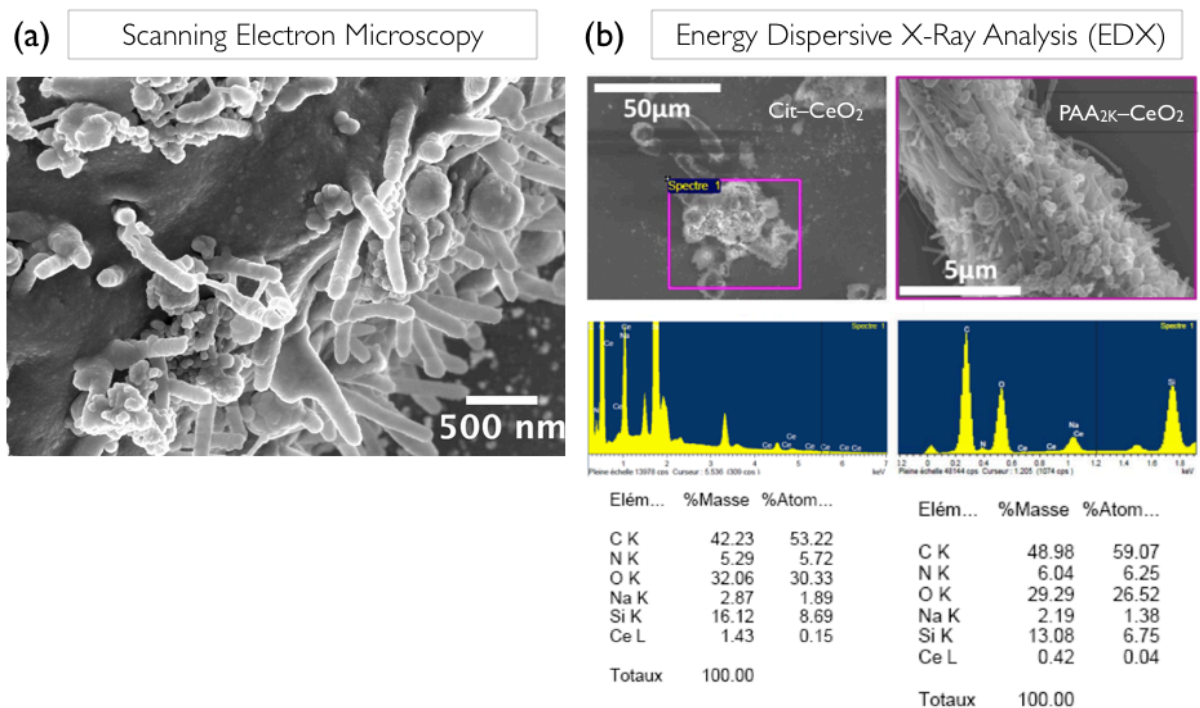

***Figure S7***: *a) SEM image of a NIH/3T3 incubated with cerium oxide nanoparticles coated with citrate ([Ce] = 1 mM) during 2 hours. b) Results of the EDX analysis for the elemental analysis.*

## S8 – MTT assays performed on NIH/3T3 mouse fibroblasts and 2139 human lymphoblastoid cells

MTT assays were performed with coated cerium oxide nanoparticles. Cells were seeded into 96-well micro-plates, and the plates were placed in an incubator overnight to allow for attachment and recovery. Cell densities were adjusted to $2 \times 10^4$ cells per well (200 µl). After 24 h, the nanoparticles were applied directly to each well using a multichannel pipette to triplicate culture wells, and cultures were incubated for 24 h at 37°C. The MTT assay depends on the cellular reduction of MTT (3-(4,5-dimethylthiazol-2-yl)-2,5-diphenyl tetrazolium bromide, Sigma-Aldrich Chemical) by the mitochondrial dehydrogenase of viable cells forming a blue formazan product that can be measured spectrophotometrically. MTT was prepared at 5 mg mL$^{-1}$ in PBS (with calcium and magnesium, Dulbecco's, PAA Laboratories) and then diluted 1 to 5 in medium without serum and without Phenol Red. After 24 h of incubation with nanoparticles, the medium was removed and 200 μl of the MTT solution was added to the microculture wells. After 4 h incubation at 37°C, the MTT solution was removed and 100 μl of 100% DMSO were added to each well to solubilize the MTT-formazan product. The absorbance at 562 nm was then measured with a microplate reader (Perkin-Elmer). Prior to the microplate UV-Vis spectrometry, MTT assays without particles were carried out with cell populations ranging from 5000 to 500000 cells and it was checked that the absorbance of DMSO solutions at 562 nm was proportional to the initial number of cells.

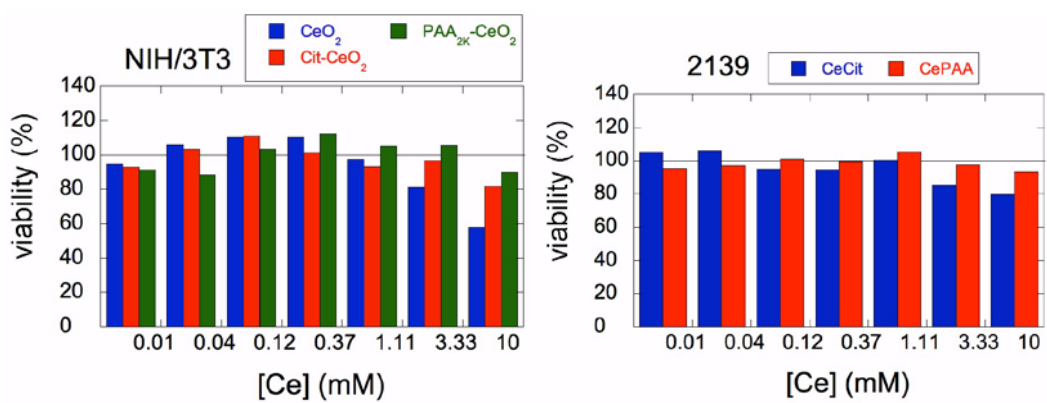

***Figure S8:*** *MTT (3-(4,5-dimethylthiazol-2-yl)-2,5-diphenyl tetrazolium bromide) viability assays conducted on NIH/3T3 fibroblasts (left) and on 2139 lymphoblasts (right) incubated with bare and coated CeO₂. The molar concentrations [Ce] were varied from 10 μM to 10 mM and the incubation time was 24 h.*

## S9 − Side and Forward Scatter Intensities as received from Flow Cytometry

As mentioned in the main text, the forward (FS) and side (SS) scatter intensities generated by the cell illumination are probing the cellular size and refractive index respectively. Figs. S9a and S9b display the SS and FS intensities (median values) of NIH/3T3 that were incubated with the coating agents adsorbed onto the nanoceria, namely sodium citrate and poly(sodium acrylate). The experimental conditions were an incubation time of 24 h, a temperature of T = 37 °C, and a concentration range comprised between 0.1 and 10 mM. These concentrations were representative of the amounts of citrates and PAA$_{2K}$ adsorbed onto nanoceria during the incubation with nanoparticles. Living and dead cells were analyzed separately. Dead cells were distinguished by their strong red fluorescence linked to the nuclear incorporation of propidium iodide. The data are compared to those of untreated cells. For both coating agents, and within the experimental uncertainties, the SS and FS signals for living and dead cells remained unchanged at all incubation doses. Together with the PI data that show no induced toxicity due to these organics, it can be concluded that the variations of the toxicity and shift in the SS scatter for citrate coated particles are related to the particles.

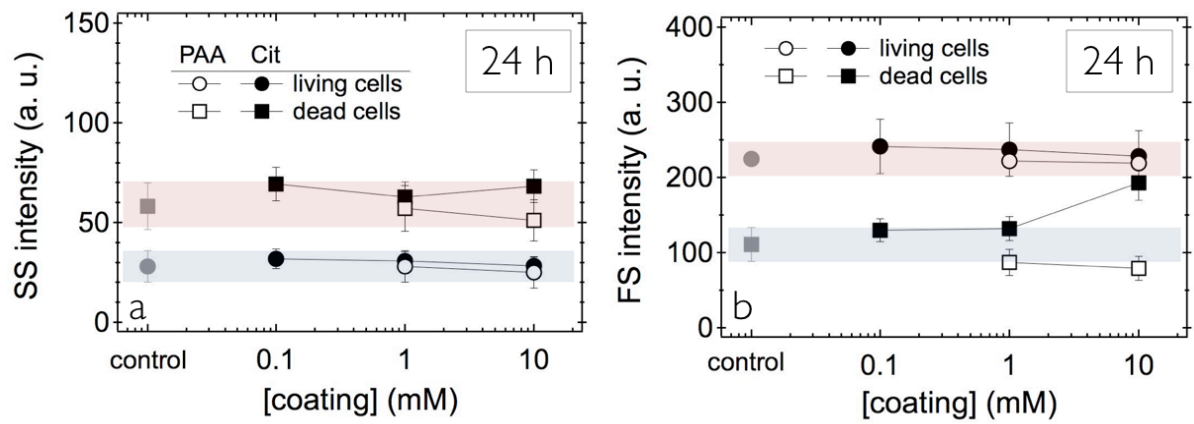

***Figure S9***: *Side (a) and forward (b) scatter intensities of NIH/3T3 incubated 24 h with poly(acrylic acid) (PAA$_{2K}$) and with citrate ligands. For the coating, the concentrations indicated in the figure are those of citrates and of acrylate monomers.*